\documentstyle[psfig]{aipproc}

\def\arcmin{\hbox{$^\prime$}}
\def\arcsec{\hbox{$^{\prime\prime}$}}
\def\degs{$^\circ$}

\begin{document}
\title{Radio Observations of the Black Hole Candidate GX~339--4}

\author{S. Corbel$^{1}$, R.P. Fender$^{2}$, P. Durouchoux$^{1}$, R.K. Sood$^{3}$, A.K. Tzioumis$^{4}$, R.E. Spencer$^{5}$ and D. Campbell-Wilson$^{6}$\\}
\address{$^{1}$CEA Saclay, DAPNIA-SAp, 91191 Gif sur Yvette Cedex, France \\
$^{2}$Astronomy Centre, University of Sussex, Falmer, Brighton BN1 9QH, UK \\
$^{3}$Australian Defence Force Academy, UNSW, Canberra, ACT 2600, Australia \\
$^{4}$Australia Telescope National Facility, CSIRO, PO Box 76, Epping 2121, NSW, Australia \\
$^{5}$University of Manchester, Jodrell Bank, Macclesfield, Cheshire SK 11 9DL, UK \\
$^{6}$School of Physics, University of Sydney, Sydney, NSW 2006, Australia \\}

\maketitle

\begin{abstract}
The black hole candidate GX~339--4 was first detected as a variable radio source by Sood \& Campbell - Wilson \cite{soo94} in May 1994 with the Molonglo Observatory Synthesis Telescope (MOST). 
Since then, several observations have been obtained with the Australian Telescope Compact Array (ATCA) in order to study the radio behavior of this source in relation to its soft and hard X-ray activity. 
We present new results of high resolution radio observations performed with the ATCA in order to study the jet-like feature observed in GX~339--4 by Fender et al \cite{fen97}. From the ATCA lightcurve at 8640 MHz, we find evidence of quenched radio emission from GX~339--4.
\end{abstract}

\section*{Introduction}
Since its discovery in 1973 with the OSO-7 satellite \cite{mar73}, GX 339--4 has been extensively studied in optical, X-rays and gamma rays. The optical counterpart was identified with a 17$^{th}$ magnitude blue star by Doxsey et al. \cite{dox79}, which was found to be very variable (15 to 20 mag.). 
Based on its bimodal spectral behavior and its fast timing properties,  GX 339--4 has been classified as a black hole candidate in a low mass system. GX 339--4 displays all three spectral modes (low state, high state and very high state) typical of black hole candidates based on the flux in the 2-10 keV band. 
Recently, M\'endez et al. \cite{men97} have found evidence for an intermediate state between the low state and the hard state. 
The X-ray spectral behavior of GX 339--4 is quite similar to that observed in the well known black hole candidate Cyg X--1.

Some anti-correlation between X-ray states and optical brightness has been reported by Motch et al. \cite{mot85}. The companion star is very faint in the off state (or very low state), brightens during the low (hard) state and is at an intermediate level during the high (soft) state. 
A brightness modulation of 14.8 hr, interpreted as the orbital period of the binary system, has been reported \cite{cal92}. But due to the strong optical emission from the accretion disk, the orbital parameters of GX 339--4 have not yet been established in order to clearly demonstrate that it is a black hole binary.
X-ray and optical quasi periodic oscillations have been observed at different frequencies, most of them during low state \cite{ste97}. Its distance estimates range from 1.3 to 4 kpc, with a favorable distance of 1.3 kpc from X-ray halo measurement \cite{pre91}.

A variable radio source was first detected by Sood \& Campbell--Wilson \cite{soo94} in May 1994 with the MOST and since then, several observations have been performed in order to study the radio behavior of this source. 
High resolution radio observations by Fender et al \cite{fen97} in July 1996 revealed the existence of a possible jet like feature issuing from the core of GX 339--4. Here we will concentrate on the observations conducted in February 1997.

\section*{Observations}

Radio observations of GX 339--4 began on May 25 1994 and are continuing at present. Data were collected from the Molonglo Observatory Synthesis Telescope (MOST) and from the Australian Telescope Compact Array (ATCA) at a total of five frequencies: 843 MHz at the MOST, 1.5, 2.4, 4.7 and 8.6 GHz at the ATCA.
The ATCA is a synthesis telescope consisting of 6 parabolic dishes, 22 m in diameter, mounted on an East-West array. 
At the time of our observations, the array was in the 6 km configuration with a maximum baseline of $\sim$ 6 km. The nominal angular resolution was $\sim$ 1 arcsec at 3.5 cm.
Amplitude calibration was performed using B1934-638, while B1646-50 was used as the phase calibrator. The source was offset from the center and the distance between the target and the phase calibrator was small (2.8 degrees). 
We used a cycle of 8 min on source and 2 min on the phase calibrator in order to remove any phase error that could introduce artifacts in the imaging process. 
The data were edited and calibrated using the software package Miriad.

\section*{Search for radio jets}

Miyamoto et al. \cite{miy91} predicted the existence of a jet in GX 339--4 in its very high state. High resolution radio observations at 3.5 cm with ATCA in July 1996 revealed a possible extension to the West of the source \cite{fen97}, which was detected at a level of $\sim$ 6 mJy.
While phase errors cannot be ruled out as the origin for this feature, it is interesting to note that just before these observations, the source's radio emission had been undetectable at an upper limit of 1 mJy from MOST. 
This is very reminiscent of the behaviour of Cygnus X-3, which undergoes periods of quenched radio emission prior to jet-producing outbursts \cite{wal96}.

\begin{figure}[h]
\centerline{\psfig{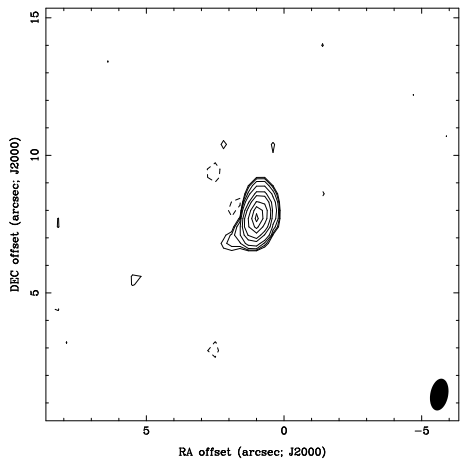}}
\caption[]{\label{gxjet} 3.5 cm map of GX 339-4 obtained with ATCA on 1997 Feb 3. Contours are -3, 3, 4, 6, 10, 20, 30, 50, 70 and 90 times r.m.s. noise of 90 $\mu$Jy beam$^{-1}$. The solid ellipse represents the synthesised beam (restored beam of 1.17\arcsec $\times$ 0.66\arcsec). The position (J2000) of the core is $\alpha$: 17 h 02 m 49.3 s and $\delta$: -48\degs 47\arcmin 31.0\arcsec}
\end{figure}

In order to confirm the existence of this feature, we performed 3 observations spaced by one week in February 1997 (3, 10 and 17) in a similar configuration. 
A very weak extension was detected in the 3.5 cm map (Figure \ref{gxjet}) of Feb 3 around the core of GX~339--4 on the opposite side of the feature observed by Fender et al. \cite{fen97}. 
The axis of the elongation is consistent with an SE-NW orientation. There is no evidence for a counterjet, that could point to a relativistic plasmoid ejection.
However, the source is compatible with a point source for the other two observations (Feb. 10 and 17). 
Due to the weakness of this feature, these observations are not enough to confirm unambiguously the presence of a jet-like feature in GX 339--4, even if the jets might have disappeared between two observations, because they are transient in nature in most X-ray binaries.
It is worth noting that all our ATCA observations were performed during an X-ray low state.\\

\section*{Radio light curve and spectra}

\begin{figure}[t]
\centerline{\psfig{file=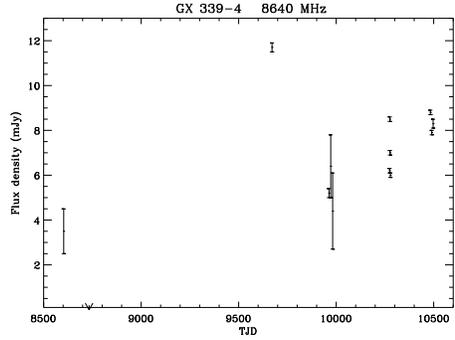,angle=0.,width=6.5cm}}
\caption[]{\label{gxlc} ATCA lightcurve of GX~339--4 at 8640 MHz}
\end{figure}

The radio lightcurve of GX~339--4 observed with ATCA at 8640 MHz is displayed in Figure \ref{gxlc}. GX~339--4 is clearly detected as a variable radio source with variations from week to week, but no flaring activity (sudden increase from an essentially quiescent level) has yet been detected, such as that seen in Cyg~X--3 \cite{wal96}, GRO J1655-40 \cite{hel95} or GRS 1915+105 \cite{rod95}. This is supported by data from the MOST (Hunstead, 1997, private communication). However, we should emphasize that the lightcurve is too sparsely sampled at present to allow us to characterize the time variability with confidence.
A slight variation at 3.5 cm ($\sim$ 10\%) is also present on a timescale of one day.

We have analysed archival data obtained with ATCA and have found that the source has been detected on December 13, 1991 at a level of $\sim$ 3 mJy. 
However, a more interesting result is the non-detection at 3.5 and 6.3 cm (with a one sigma upper limit of 0.2 mJy) on April 21, 1992, at a time when the source was very weak in hard X-rays. 
This is very similar again to the quenched radio emission observed in Cyg X-3. This is the first time radio emission was not detected from GX~339--4 at the level of this strong upper limit.\\

\begin{figure}[b]
\centerline{\psfig{file=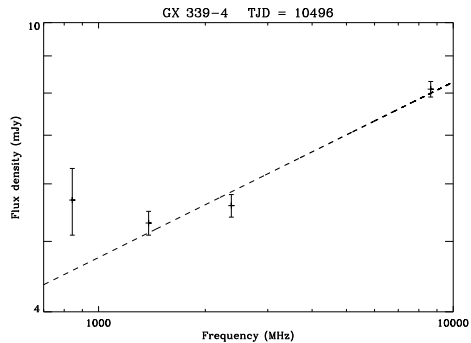,angle=0.,width=6.5cm}}
\caption[]{\label{gxspec} Radio spectra of GX 339-4 from 843 to 8640 MHz. The dotted line indicates the best fit to the ATCA datapoints.}
\end{figure}

The radio spectrum of GX~339--4 obtained on 1997 February 17 from 843 MHz (36 cm) to 8640 MHz (3 cm) is presented in Figure \ref{gxspec}. Observations at 843 MHz were performed with the MOST.
Using the convention S($\nu$) $\propto \nu ^{\alpha}$, the spectrum could be fitted with a single powerlaw with an average spectral index of + 0.24 if we excluded the MOST datapoint, but the latter seems to indicate a curvature in the spectrum and could suggest the presence of another emission component. 
The inverted spectrum component indicates a compact core which is partially optically thick, similar to previous observation\cite{fen97}.

\section*{Discussion}

Among the radio emitting X-ray binaries, GX~339--4 is a peculiar source. Its radio behavior is quite similar to that of Cyg~X--1 (variable and flat radio spectra), but with very different temporal hard X-ray activity.
Strong outbursts have been detected with BATSE \cite{rob97} in hard X-rays from GX~339--4.
The superluminal source, GRO J1655-40, exhibits a hard X-ray behavior ``similar'' to that of GX~339--4, with the main difference being that GX~339--4 doesn't show radio flaring and superluminal radio ejection.
Recently, Mioduszewski et al. \cite{mio97} have shown some evidence that Cyg X-3 might also be a superluminal source.
The detection of quenched radio emission (if confirmed) from GX~339--4 prior to radio ejection could be a missing link between these sources. The long term radio behavior of GX 339--4 could link the jet sources (and transient radio sources) to Cyg X-1 (persistent radio source) and Cyg X-3.
All our radio observations have been performed during the low state of the source, and a more complete coverage of the various X-ray states should provide new clues for the understanding of GX~339--4.

ACKNOWLEDGMENTS:
\noindent
The authors would like to thank Dick Hunstead and Mike Nowak for useful discussions and Vince McIntyre and Marc Elmouttie for helpful assistance during the ATCA observations.
S. Corbel acknowledges support from Le Comit\'e National Fran\c cais  d'Astronomie and the local organizers of the Compton symposium.
The Australia Telescope is funded by the Commonwealth of Australia for operation as a National Facility managed by CSIRO. The MOST is supported by the Australian Research Council and the Science Foundation for Physics within the University of Sydney. 

\end{document}